\documentclass{llncs}

\def\half{\textstyle{1 \over 2}}
\def\01{\{0,1\}}
\def\x{\times}
\def\ox{}
\def\xor{+}
\def\AND{}
\def\Q{\mbox{\sf Q}}
\def\ket#1{\mbox{$| #1 \rangle$}}
\def\state#1{\mbox{state$( #1 )$}}
\def\sqm#1{\textstyle{\sqrt{#1}}}
\def\half{\textstyle{1 \over 2}}

\def\ee{\vspace*{2mm}}
\def\eee{\vspace*{5mm}}
\def\imag{\mbox{i}}
\def\loud#1{\noindent{\bf #1 }}
\def\a{\alpha}
\def\b{\beta}
\def\e{\varepsilon}
\def\IP{{\it IP\/}}
\def\mIP{\mbox{\it IP\/}}
\def\Q{Q_0^{\ast}}
\def\C{C_0^{\ast}}
\def\Qe{Q_{\e}^{\ast}}
\def\Ce{C_{\e}^{\ast}}

\begin{document}

\title{Quantum Entanglement and the Communication Complexity of 
the Inner Product Function}

\author{
Richard Cleve\inst{1}%
\thanks{Research initiated while visiting the Universit\'e de Montr\'eal 
and supported in part by Canada's NSERC.}
\and 
Wim van Dam\inst{2} 
\and 
Michael Nielsen\inst{3}%
\and 
Alain Tapp\inst{4}%
\thanks{Research supported in part by Canada's NSERC.} 
}

\institute{
University of Calgary%
\thanks{Department of Computer Science, University of Calgary, Calgary, 
Alberta, Canada T2N 1N4.
E-mail: {\tt cleve@cpsc.ucalgary.ca}.}
\and
University of Oxford and CWI, Amsterdam%
\thanks{Clarendon Laboratory, Department of Physics, University of Oxford, 
Parks Road, Oxford OX1 3PU, U.K.
E-mail: {\tt wimvdam@mildred.physics.ox.ac.uk}.} 
\and
Los Alamos National Laboratory and University of New Mexico%
\thanks{T-6 Theoretical Astrophysics, Los Alamos National Laboratory, U.S.A.
E-mail: {\tt mnielsen@tangelo.phys.unm.edu}.}
\and
Universit\'e de Montr\'eal%
\thanks{D\'epartement IRO, C.P. 6128, Succursale Centre-Ville, Montr\'eal, 
Qu\'ebec, Canada H3C 3J7.
E-mail: {\tt tappa@iro.umontreal.ca}.}}

\maketitle

\begin{abstract}
We consider the communication complexity of the binary inner product 
function in a variation of the two-party scenario where the parties 
have an {\it a priori\/} supply of particles in an entangled quantum 
state.
We prove linear lower bounds for both exact protocols, as well as 
for protocols that determine the answer with bounded-error probability.
Our proofs employ a novel kind of ``quantum'' reduction from a quantum 
information theory problem to the problem of computing the inner product.
The communication required for the former problem can then be bounded 
by an application of Holevo's theorem.
We also give a specific example of a probabilistic scenario 
where entanglement reduces the communication complexity of the inner 
product function by one bit.
\end{abstract}

\section{Introduction and Summary of Results}

The {\em communication complexity\/} of a function 
$f : \01^n \x \01^n \rightarrow \01$ is defined as the minimum 
amount of communication necessary among two parties, conventionally 
referred to as Alice and Bob, in order for, say, Bob to acquire the 
value of $f(x,y)$, where, initially, Alice is given $x$ and Bob is 
given $y$.
This scenario was introduced by Yao \cite{Yao79} and has been widely 
studied (see \cite{KN} for a survey).
There are a number of technical choices in the model, such as: 
whether the communication cost is taken as the worst-case $(x,y)$, 
or the average-case $(x,y)$ with respect to some probability
distribution; 
whether the protocols are deterministic or probabilistic (and, for 
probabilistic protocols, whether the parties have independent random 
sources or a shared random source); and, what correctness probability 
is required.

The communication complexity of the {\it inner product modulo two (IP)} 
function 
\begin{eqnarray}
\mIP(x,y) &=& x_1 \AND y_1 \xor x_2 \AND y_2 \xor \cdots \xor 
x_n \AND y_n \bmod 2
\label{IP}
\end{eqnarray}
is fairly well understood in the above ``classical'' models.
For worst-case inputs and deterministic errorless protocols, 
the communication complexity is $n$ and, for randomized protocols 
(with either an independent or a shared random source), uniformly 
distributed or worst-case inputs, and with error probability 
${1 \over 2} - \delta$ required, the communication complexity 
is $n - O(\log (1 / \delta))$ \cite{CG88} (see also \cite{KN}).

In 1993, Yao \cite{Yao93} introduced a variation of the above classical 
communication complexity scenarios, where the parties communicate with 
{\it qubits\/}, rather than with bits.
Protocols in this model are at least as powerful as probabilistic 
protocols with independent random sources.
Kremer \cite{Kremer} showed that, in this model, the communication 
complexity of \IP\ is $\Omega(n)$, whenever the required correctness 
probability is $1 - \varepsilon$ for a constant $0 \leq \e < {1 \over 2}$ 
(Kremer attributes the proof methodology to Yao).

Cleve and Buhrman \cite{CB} (see also \cite{BCD}) introduced another 
variation of the classical communication complexity scenario that also 
involves quantum information, but in a different way.
In this model, Alice and Bob have an initial supply of particles in an 
entangled quantum state, such as Einstein-Podolsky-Rosen (EPR) pairs, 
but the communication is still in terms of classical bits.
They showed that the entanglement enables the communication for a 
specific problem to be reduced by one bit.
Any protocol in Yao's qubit model can be simulated by a protocol in 
this entanglement model with at most a factor two increase in 
communication: each qubit can be ``teleported'' \cite{BBCJPW} by 
sending two classical bits in conjunction with an EPR pair of 
entanglement.
On the other hand, we are aware of no similar simulation of protocols 
in the entanglement model by protocols in the qubit model, and, thus, the 
entanglement model is potentially stronger.

In this paper, we consider the communication complexity of \IP\ in two 
scenarios: with prior entanglement and qubit communication; and with 
prior entanglement and classical bit communication.
As far as we know, the proof methodology of the lower bound in the qubit 
communication model without prior entanglement \cite{Kremer} does not 
carry over to either of these two models.
Nevertheless, we show $\Omega(n)$ lower bounds in these models.

To state our lower bounds more precisely, we introduce the following 
notation.
Let $f : \01^n \x \01^n \rightarrow \01$ be a communication problem, and 
$0 \le \epsilon < \half$.
Let $\Qe(f)$ denote the communication complexity of $f$ in terms of 
{\em qubits}, where quantum entanglement is available and the requirement 
is that Bob determines the correct answer with probability at least 
$1 - \e$ (the $\ast$ superscript is intended to highlight the fact that prior 
entanglement is available).
Also, let $\Ce(f)$ denote the corresponding communication complexity of 
$f$ in the scenario where the communication is in terms of {\em bits} (again, 
quantum entanglement is available and Bob is required to determine the 
correct answer with probability at least $1 - \e$).
When $\e = 0$, we refer to the protocols as {\em exact}, and, when $\e > 0$, 
we refer to them as {\em bounded-error} protocols.
With this notation, our results are:
\begin{eqnarray}
\Q(\mIP)  & = & \lceil n/2 \rceil \label{Q} \\
\Qe(\mIP) & \ge & \half(1 - 2 \e)^2 n - \half \label{Qe} \\
\C(\mIP)  & = & n \label{C} \\
\Ce(\mIP) & \ge & \max(\half(1 - 2 \e)^2,(1 - 2 \e)^4) n - \half \label{Ce} 
\end{eqnarray}
Note that all the lower bounds are $\Omega(n)$ whenever $\e$ is held 
constant.
Also, these results subsume the lower bounds in \cite{Kremer}, since the 
qubit model defined by Yao \cite{Yao93} differs from the bounded-error 
qubit model defined above only in that it does not permit a prior 
entanglement. 

Our lower bound proofs employ a novel kind of ``quantum'' reduction 
between protocols, which reduces the problem of communicating, say, 
$n$ bits of information to the \IP\ problem.
It is noteworthy that, in classical terms, it can be shown that there is 
no such reduction between the two problems.
The appropriate cost associated with communicating $n$ bits is then 
lower-bounded by the following nonstandard consequence of Holevo's 
theorem.\ee

\loud{Theorem 1:}{\sl In order for Alice to convey $n$ bits of information 
to Bob, where quantum entanglement is available and qubit communication in 
either direction is permitted, Alice must send Bob at least 
$\lceil n/2 \rceil$ qubits.
This holds regardless of the prior entanglement and the qubit communication 
from Bob to Alice.
More generally, for Bob to obtain $m$ bits of mutual information with 
respect to Alice's $n$ bits, Alice must send at least $\lceil m/2 \rceil$ 
qubits.}\ee

A slight generalization of Theorem 1 is described and proven in 
the Appendix.

It should be noted that, since quantum information subsumes classical 
information, our results also represent new proofs of nontrivial lower 
bounds on the {\em classical\/} communication complexity of \IP, and our 
methodology is fundamentally different from those previously used for 
classical lower bounds.

Finally, with respect to the question of whether quantum entanglement 
can {\em ever} be advantageous for protocols computing \IP, we present 
a curious probabilistic scenario with $n=2$ where prior entanglement 
enables one bit of communication to be saved.

\section{Bounds for Exact Qubit Protocols}

In this section, we consider exact qubit protocols computing \IP, 
and prove Eq.~(\ref{Q}).
Note that the upper bound follows from so-called ``superdense coding'' 
\cite{BW92}: by sending $\lceil n/2 \rceil$ qubits in conjunction with 
$\lceil n/2 \rceil$ EPR pairs, Alice can transmit her $n$ classical bits 
of input to Bob, enabling him to evaluate \IP.
For the lower bound, we consider an arbitrary exact qubit protocol 
that computes \IP, and convert it (in two stages) to a protocol for which 
Theorem 1 applies.

For convenience, we use the following notation.
If an $m$-qubit protocol consists of $m_1$ qubits from Alice to Bob and 
$m_2$ qubits from Bob to Alice then we refer to the protocol as an 
$(m_1,m_2)$-qubit protocol.

\subsection{Converting Exact Protocols into Clean Form}

A {\em clean protocol\/} is a special kind of qubit protocol that follows 
the general spirit of the reversible programming paradigm in a quantum 
setting.
Namely, one in which all qubits incur no net change, except for one, 
which contains the answer.

In general, the initial state of a qubit protocol is of the form 
\begin{equation}
\underbrace{\ket{y_1,\ldots,y_n} \ox \ket{0,\ldots,0} \ox |\Phi}%
_{\mbox{\small Bob's qubits}}
\underbrace{\!{}_{BA}\rangle \ox \ket{x_1,\ldots,x_n} \ox 
\ket{0,\ldots,0}}%
_{\mbox{\small Alice's qubits}},
\end{equation}
where $\ket{\Phi_{\!BA}}$ is the state of the entangled qubits shared by 
Alice and Bob, and the $\ket{0,\ldots,0}$ states can be regarded as 
``ancillas''.
At each turn, a player performs some transformation (which, without loss 
of generality, can be assumed to be unitary) on all the qubits in his/her 
possession and then sends a subset of these qubits to the other player.
Note that, due to the communication, the qubits possessed by each player 
varies during the execution of the protocol.
At the end of the protocol, Bob measures one of his qubits which is 
designated as his {\em output}.

We say that a protocol which exactly computes a function $f(x,y)$ is 
{\em clean\/} if, when executed on the initial state 
\begin{equation}
\ket{z} \ox \ket{y_1,\ldots,y_n} \ox \ket{0,\ldots,0} \ox \ket{\Phi_{BA}}  
\ox \ket{x_1,\ldots,x_n} \ox \ket{0,\ldots,0}, 
\label{initial}
\end{equation}
results in the final state 
\begin{equation}
\ket{z \xor f(x,y)} \ox \ket{y_1,\ldots,y_n}  
\ox \ket{0,\ldots,0} \ox \ket{\Phi_{BA}} \ox \ket{x_1, \ldots, x_n} 
\ox \ket{0,\ldots,0}
\label{final}
\end{equation}
(where the addition is mod 2).
The ``input'', the ancilla, and initial entangled qubits will typically 
change states during the execution of the protocol, but they are reset 
to their initial values at the end of the protocol.

It is straightforward to transform an exact $(m_1,m_2)$-qubit protocol 
into a clean $(m_1+m_2,m_1+m_2)$-qubit protocol that computes the same 
function.
To reset the bits of the input, the ancilla, and the initial entanglement,  
the protocol is run once, except the output is not measured, but recorded 
and then the protocol is run in the {\em backwards} direction to 
``undo the effects of the computation''.
The output is recorded on a {\em new} qubit of Bob (with initial 
state $\ket{z}$) which is control-negated with the output qubit of Bob 
(that is in the state $\ket{f(x,y)}$) as the control.
Note that, for each qubit that Bob sends to Alice when the protocol 
is run forwards, Alice sends the qubit to Bob when run in the backwards 
direction.
Running the protocol backwards resets all the qubits---except Bob's 
new one---to their original states.
The result is an $(m_1+m_2,m_1+m_2)$-qubit protocol that maps state 
(\ref{initial}) to state (\ref{final}).

\subsection{Reduction from Communication Problems}

We now show how to transform a clean $(m_1+m_2,m_1+m_2)$-qubit 
protocol that exactly computes \IP\ for inputs of size $n$, to an 
$(m_1+m_2,m_1+m_2)$-qubit protocol that transmits $n$ bits of information 
from Alice to Bob.
This is accomplished in four stages:  
\begin{enumerate}
\item
Bob initializes his qubits indicated in Eq.~(\ref{initial}) 
with $z = 1$ and $y_1 = \cdots = y_n = 0$.
\item
Bob performs a Hadamard transformation on each of his first $n+1$ qubits.
\item 
Alice and Bob execute the clean protocol for the inner product function.
\item
Bob again performs a Hadamard transformation on each of his first 
$n+1$ qubits.
\end{enumerate}
Let $\ket{B_i}$ denote the state of Bob's first $n+1$ qubits after the 
$i^{\mbox{\scriptsize th}}$ stage.
Then 
\begin{eqnarray}
\ket{B_1} &=& \ket{1} \ox \ket{0,\ldots,0}  \label{st1} \\
\ket{B_2} &=& {\textstyle{1 \over \sqrt{2^{n+1}}}}
\sum_{a,b_1,\ldots,b_n \in \01}
(-1)^a\ket{a} \ox \ket{b_1,\ldots,b_n} \\
\ket{B_3} &=& {\textstyle{1 \over \sqrt{2^{n+1}}}}
\sum_{a,b_1,\ldots,b_n \in \01} 
(-1)^a\ket{a + b_1 x_1 + \cdots + b_n x_n} \ox \ket{b_1,\ldots,b_n} \nonumber \\
&=& {\textstyle{1 \over \sqrt{2^{n+1}}}}
\sum_{c,b_1,\ldots,b_n \in \01} 
(-1)^{c + b_1 \AND x_1 \xor \cdots \xor b_n \AND x_n}
\ket{c} \ox \ket{b_1,\ldots,b_n} \label{st3} \\
\ket{B_4} &=& \ket{1} \ox \ket{x_1,\ldots,x_n}, \label{st4}
\end{eqnarray}
where, in Eq.~(\ref{st3}), the substitution 
$c = a + b_1 x_1 + \cdots + b_n x_n$ has been made (and arithmetic over 
bits is taken mod 2).
The above transformation was inspired by the reading of \cite{TS} (see 
also \cite{BV}).

Since the above protocol conveys $n$ bits of information (namely, 
$x_1,\ldots,x_n$) from Alice to Bob, by Theorem 1, we have 
$m_1+m_2 \ge n/2$.
Since this protocol can be constructed from an arbitrary 
exact $(m_1,m_2)$-qubit protocol for \IP, this establishes the 
lower bound of Eq.~(\ref{Q}).

Note that, classically, no such reduction is possible.
For example, if a clean protocol for \IP\ is executed in any classical 
context, it can never yield more than one bit of information to Bob 
(whereas, in this quantum context, it yields $n$ bits of information 
to Bob).

\section{Lower Bounds for Bounded-Error Qubit Protocols}

In this section we consider bounded-error qubit protocols for \IP, and 
prove Eq.~(\ref{Qe}).
Assume that some qubit protocol $P$ computes \IP\ correctly with 
probability at least $1 - \e$, where $0 < \e < \half$.
Since $P$ is not exact, the constructions from the previous section do 
not work exactly.
We analyze the extent by which they err.  

First, the construction of Section 2.1 will not 
produce a protocol in clean form; however, it will result in a protocol 
which {\em approximates} an exact clean protocol (this type of construction 
was previously carried out in a different context by Bennett {\em et al.} 
\cite{BBBV97}).

Denote the initial state as 
\begin{equation}
\ket{y_1,\dots,y_n} \ox \ket{0,\dots,0} \ox \ket{\Phi_{BA}} 
\ox \ket{x_1,\dots,x_n} \ox \ket{0,\dots,0}.
\label{start}
\end{equation}
Also, assume that, in protocol $P$, Bob never changes the state 
of his input qubits $\ket{y_1,\ldots,y_n}$ (so the first $n$ qubits 
never change).
This is always possible, since he can copy $y_1,\ldots,y_n$ into his 
ancilla qubits at the beginning.
After executing $P$ until just before the measurement occurs, the 
state of the qubits must be of the form
\begin{equation}
\a \ket{y_1,\ldots,y_n} \ox \ket{x \cdot y} \ox \ket{J} + 
\b \ket{y_1,\ldots,y_n} \ox \ket{\overline{x \cdot y}} \ox \ket{K}, 
\end{equation}
where $|\a|^2 \ge (1-\e)$ and $|\b|^2 \le \e$.
In the above, the $n+1^{\mbox{\scriptsize st}}$ qubit is the designated 
output, $x \cdot y$ denotes the inner product of $x$ and $y$, and 
$\overline{x \cdot y}$ denotes the negation of this inner product.
In general, $\a$, $\b$, $\ket{J}$, and $\ket{K}$ may depend on $x$ and $y$.

Now, suppose that the procedure described in Section 2.1 for producing 
a clean protocol in the exact case is carried out for $P$.
Since, in general, the answer qubit is not in the state 
$\ket{x \cdot y}$---or even in a pure basis state---this does not 
produce the final state 
\begin{equation}
\ket{z + x \cdot y} \ox \ket{y_1,\ldots,y_n} \ox \ket{0,\ldots,0} \ox 
\ket{\Phi_{BA}} \ox \ket{x_1,\ldots,x_n} \ox \ket{0,\ldots,0}.
\label{exact}
\end{equation}
However, let us consider the state that is produced instead.
After introducing the {\em new} qubit, initialized in basis state $\ket{z}$, 
and applying $P$, the state is 
\begin{equation}
\ket{z} \ox 
\left(\a \ket{y_1,\ldots,y_n} \ox \ket{x \cdot y} \ox \ket{J} + 
\b \ket{y_1,\ldots,y_n} \ox \ket{\overline{x \cdot y}} \ox \ket{K}\right).
\end{equation}
After applying the controlled-NOT gate, the state is 
\begin{eqnarray}
\lefteqn{\a \ket{z+x \cdot y} \ox \ket{y_1,\ldots,y_n} \ox 
\ket{x \cdot y} \ox \ket{J} + 
\b \ket{z + \overline{x \cdot y}} \ox \ket{y_1,\ldots,y_n} \ox 
\ket{\overline{x \cdot y}} \ox \ket{K}} & & \nonumber \vspace*{1mm} \\
& = & 
\a \ket{z+x \cdot y} \ox \ket{y_1,\ldots,y_n} \ox 
\ket{x \cdot y} \ox \ket{J} + 
\b \ket{z + x \cdot y} \ox \ket{y_1,\ldots,y_n} \ox 
\ket{\overline{x \cdot y}} \ox \ket{K} \nonumber \\ 
& & - \b \ket{z + x \cdot y} \ox \ket{y_1,\ldots,y_n} \ox 
\ket{\overline{x \cdot y}} \ox \ket{K} +
\b \ket{z + \overline{x \cdot y}} \ox \ket{y_1,\ldots,y_n} \ox 
\ket{\overline{x \cdot y}} \ox \ket{K} \nonumber \vspace*{1mm} \\
& = & 
\ket{z + x \cdot y} \ox \left(
\a \ket{y_1,\ldots,y_n} \ox \ket{x \cdot y} \ox \ket{J} + 
\b \ket{y_1,\ldots,y_n} \ox \ket{\overline{x \cdot y}} \ox \ket{K}
\right) \nonumber \\
& & + \sqrt{2}\beta\left(
\textstyle{1 \over \sqrt{2}} \ket{z + \overline{x \cdot y}} - 
\textstyle{1 \over \sqrt{2}} \ket{z + x \cdot y}\right) 
\ox \ket{y_1,\ldots,y_n} \ox \ket{\overline{x \cdot y}} \ox \ket{K}.
\end{eqnarray}
Finally, after applying $P$ in reverse to this state, the final state is 
\begin{equation}
\ket{z + x \cdot y} \ox \ket{y_1,\ldots,y_n} \ox \ket{0,\ldots,0} \ox 
\ket{\Phi_{BA}} \ox \ket{x_1,\ldots,x_n} \ox \ket{0,\ldots,0}
+ \sqrt{2}\beta\ket{M_{x,y,z}}, 
\label{finish}
\end{equation}
where 
\begin{equation}
\ket{M_{x,y,z}} = \left(
\textstyle{1 \over \sqrt{2}} \ket{z + \overline{x \cdot y}} - 
\textstyle{1 \over \sqrt{2}} \ket{z + x \cdot y}\right) 
\ox P^{\mbox{\scriptsize \dag}} 
\ket{y_1,\ldots,y_n} \ox \ket{\overline{x \cdot y}} \ox \ket{K}.
\end{equation}

Note that the vector $\sqrt{2}\beta\ket{M_{x,y,z}}$ is the difference 
between what an exact protocol would produce (state (\ref{exact})) and 
what is obtained by using the inexact (probabilistic) protocol $P$ 
(state (\ref{finish})).
There are some useful properties of the $\ket{M_{x,y,z}}$ states.
First, as $y \in \01^n$ varies, the states $\ket{M_{x,y,z}}$ are 
orthonormal, since $\ket{y_1,\ldots,y_n}$ is a factor in each such 
state (this is where the fact that Bob does not change his input 
qubits is used).
Also, $\ket{M_{x,y,0}} = - \ket{M_{x,y,1}}$, since only the 
$({1 \over \sqrt{2}} \ket{z + \overline{x \cdot y}} - 
{1 \over \sqrt{2}} \ket{z + x \cdot y})$ factor in each such state 
depends on $z$.

Call the above protocol $\tilde{P}$.
Now, apply the four stage reduction in Section 2.2, with $\tilde{P}$ 
in place of an exact clean protocol.
The {\em difference} between the state produced by using $\tilde{P}$ and 
using an exact clean protocol first occurs after the third stage and is 
\begin{eqnarray}
\lefteqn{
{\textstyle{1 \over \sqrt{2^{n+1}}}} 
\sum_{y_1,\ldots,y_n,z \in \01} (-1)^z \sqrt{2} \b_y \ket{M_{x,y,z}} 
} & & \nonumber \\
& = & 
{\textstyle{1 \over \sqrt{2^{n+1}}}} 
\sum_{y_1,\ldots,y_n \in \01} \sqrt{2} \b_y \left(
\ket{M_{x,y,0}} - \ket{M_{x,y,1}}
\right) \nonumber \\
& = & 
{\textstyle{2 \over \sqrt{2^n}}} 
\sum_{y_1,\ldots,y_n \in \01} \b_y \ket{M_{x,y,0}},
\end{eqnarray}
which has magnitude bounded above by 
$2\sqrt{\e}$, 
since, for each $y \in \01^n$, \mbox{$|\b_y|^2 \le \e$}, and 
the $\ket{M_{x,y,0}}$ states are orthonormal.
Also, the magnitude of this difference does not change when the Hadamard 
transform in the fourth stage is applied.
Thus, the final state is within Euclidean distance 
$2\sqrt{\e}$ from 
\begin{equation}
\ket{1} \ox \ket{x_1,\dots,x_n} \ox \ket{0,\dots,0} \ox \ket{\Phi_{BA}} 
\ox \ket{x_1,\dots,x_n} \ox \ket{0,\dots,0}.
\label{correct}
\end{equation}
Consider the angle $\theta$ between this final state and (\ref{correct}).
It satisfies $\sin^2\theta + (1- \cos\theta)^2 \le 4 \e$, from which 
it follows that $\cos\theta \ge 1-2 \e$.
Therefore, if Bob measures his first $n+1$ qubits in the standard 
basis, the probability of obtaining $\ket{1,x_1,\dots,x_n}$ 
is $\cos^2\theta \ge (1-2\e)^2$.

Now, suppose that $x_1,\ldots,x_n$ are uniformly distributed.
Then Fano's inequality (see, for example, \cite{CT91}) implies that 
Bob's measurement causes his uncertainty about $x_1,\ldots,x_n$ to 
drop from $n$ bits to less than $(1-(1-2\e)^2) n + h((1-2 \e)^2)$ bits, 
where $h(x) = -x \log x - (1-x)\log(1-x)$ is the binary entropy 
function.
Thus, the mutual information between the result of Bob's measurement 
and $(x_1,\ldots,x_n)$ is at least 
$(1-2\e)^2 n - h((1-2\e)^2) \ge (1-2\e)^2 n - 1$ bits.
By Theorem 1, the communication from Alice to Bob is at least 
$\half(1-2\e)^2n - \half$ qubits, which establishes Eq.~(\ref{Qe}).

\section{Lower Bounds for Bit Protocols}

In this section, we consider exact and bounded-error bit protocols for 
\IP, and prove Eqs.~(\ref{C}) and (\ref{Ce}).

Recall that any $m$-qubit protocol can be simulated by a $2m$-bit 
protocol using teleportation \cite{BBCJPW} (employing EPR pairs of 
entanglement).
Also, if the communication pattern in an $m$-bit protocol is such that 
an even number of bits is always sent during each party's turn then 
it can be simulated by an $m/2$-qubit protocol by superdense 
coding \cite{BW92} (which also employs EPR pairs).
However, this latter simulation technique cannot, in general, be applied 
directly, especially for protocols where the parties take turns 
sending single bits.

We can nevertheless obtain a slightly weaker simulation of bit protocols 
by qubit protocols for \IP\ that is sufficient for our purposes.
The result is that, given any $m$-bit protocol for $\mIP_n$ (that is, \IP\ 
instances of size $n$), one can construct an $m$-qubit protocol for 
$\mIP_{2n}$.
This is accomplished by interleaving two executions of the bit protocol 
for $\mIP_n$ to compute two independent instances of inner products of 
size $n$.
We make two observations.
First, by taking the sum (mod 2) of the two results, one obtains an 
inner product of size $2n$.
Second, due to the interleaving, an even number of bits is sent at each turn, 
so that the above superdense coding technique can be applied, yielding a 
$(2m)/2 = m$-qubit protocol for $\mIP_{2n}$.
Now, Eq.~(\ref{Q}) implies $m \ge n$, which establishes the lower 
bound of Eq.~(\ref{C}) (and the upper bound is trivial).

If the same technique is applied to any $m$-bit protocol computing $\mIP_n$ 
with probability $1-\e$, one obtains an $m$-qubit protocol that 
computes $\mIP_{2n}$ with probability 
$(1 - \e)^2 + \e^2 = 1 - 2\e(1-\e)$.
Applying Eq.~(\ref{Qe}) here, with $2n$ replacing $n$ and 
$2\e(1-\e)$ replacing $\e$, yields $m \ge (1-2\e)^4 n - \half$.
For $\e > \frac{2-\sqrt{2}}{4} = 0.146\ldots$, a better bound is obtained 
by simply noting that $\Ce \ge \Qe$ (since qubits can always be used in 
place of bits), and applying Eq.~(\ref{Qe}). 
This establishes Eq.~(\ref{Ce}).

%
\section{An Instance where Prior Entanglement is Beneficial}

Here we will show that in spite of the preceding results,
it is still possible that a protocol which uses prior entanglement
outperforms all possible classical protocols. 
This improvement is done in the probabilistic sense where we look
at the number of communication bits required to reach a certain 
reliability threshold for the \IP\ function. 
This is done in the following setting. 

Both Alice and Bob have a 2 bit vector $x_1 x_2$ and
$y_1 y_2$, for which they want to calculate
the inner product modulo 2:
\begin{eqnarray}
f(x,y) & = & x_1 \AND y_1 \xor x_2 \AND y_2 \bmod 2
\end{eqnarray}
with a correctness-probability of at least $4 \over 5$.
It will be shown that
with entanglement Alice and Bob can reach this ratio 
with 2 bits of communication, whereas without entanglement 
3 bits are necessary to obtain this success-ratio.

\subsection{A Two-Bit Protocol with Prior Entanglement}

Initially Alice and Bob share a joint random coin and
an EPR-like pair of qubits $Q_A$ and $Q_B$:
\begin{eqnarray}
\state{Q_A Q_B} & = & 
\textstyle{{1 \over \sqrt{2}}(\ket{00} + \ket{11}})
\end{eqnarray}
With these attributes the protocol goes as follows.

First Alice and Bob determine by a joint random coin flip\footnote{ 
Because a joint random coin flip can be simulated with an EPR-pair,
we can also assume that Alice and Bob start the protocol 
with two shared EPR-pairs and no random coins.}
who is going to be the `sender' and the `receiver'
in the protocol. 
(We continue the description of the protocol by assuming that 
Alice is the sender and that Bob is the receiver.)
After this, Alice (the sender) applies the rotation $A_{x_1 x_2}$ on
her part of the entangled pair and measures 
this qubit $Q_A$ in the standard basis. The result $m_A$ 
of this measurement is then sent to Bob (the receiver)
who continues the protocol. 

If Bob has the input string `00', he knows with certainty 
that the outcome of the function $f(x,y)$ is zero and hence 
he concludes the protocol by  
sending the bit $0$ to Alice. 
Otherwise, Bob performs the rotation 
$B_{y_1 y_2}$ on his part of the entangled pair $Q_B$ 
and measure it in the standard basis 
yielding the value $m_B$.
Now Bob finishes the protocol by sending to Alice the 
bit $m_A \xor m_B \bmod 2$.

Using the rotations shown below and bearing in mind the randomization
process in the beginning of the protocol with the joint coin flip, 
this will be a protocol that uses only 2 bits of classical 
communication and that gives the correct value of $f(x,y)$ with a 
probability of at least ${4 \over 5}$ for every possible combination 
of $x_1 x_2$ and $y_1 y_2$. 

The unitary transformations used by the sender in the protocol are:
\begin{eqnarray}
\begin{array}{ll}
A_{00} = \left({\begin{array}{rr}
\sqm{2 \over 5} & \ \ -\imag \sqm{3 \over 5} \vspace*{2mm} \\
-\imag \sqm{3 \over 5} & \sqm{2 \over 5}
\end{array}}\right)
& 
A_{01} = \left({\begin{array}{cc}
\sqm{4 \over 5} & \sqm{3 \over 16} + \imag \sqm{1 \over 80} \vspace*{2mm} 
\vspace*{2mm} \\
-\sqm{3 \over 16} + \imag \sqm{1 \over 80} & \sqm{4 \over 5}
\end{array}}\right)
\vspace*{9mm} \\
A_{10} = \left({\begin{array}{cc}
\sqm{4 \over 5} & -\sqm{3 \over 16} + \imag \sqm{1 \over 80} \vspace*{2mm}
\\
\sqm{3 \over 16} + \imag \sqm{1 \over 80} & \sqm{4 \over 5}
\end{array}}\right) 
&
A_{11} = \left({\begin{array}{rr}
\sqm{1 \over 5} & \ \ \imag \sqm{4 \over 5} \vspace*{2mm} \\
\imag \sqm{4 \over 5} & \sqm{1 \over 5} 
\end{array}}\right) ,
\end{array}
\end{eqnarray}
whereas the receiver uses one of the three rotations:
\begin{eqnarray} \label{Brot}
\begin{array}{ll}
B_{01}  =  \left({\begin{array}{cc}
\sqm{3 \over 5} & -\half + \imag \sqm{3 \over 20} \vspace*{2mm} \\
-\half -\imag \sqm{3 \over 20} & -\sqm{3 \over 5}
\end{array}}\right)
&
B_{10} = \left({\begin{array}{cc}
\sqm{3 \over 5} & \half + \imag \sqm{3 \over 20} \vspace*{2mm} \\
-\half + \imag \sqm{3 \over 20} & \sqm{3 \over 5}
\end{array}}\right)
\vspace*{3mm} \\
B_{11} = \left({\begin{array}{cc}
0 & 1 \\
1 & 0
\end{array}}\right) .
&
\end{array}
\end{eqnarray}
The matrices were found by using an optimization
program that suggested certain numerical values. 
A closer examination of these values revealed the above
analytical expressions.

\subsection{No Two-Bit Classical Probabilistic Protocol Exists}

Take the probability distribution $\pi$ on the input strings $x$ and 
$y$, defined by:
\begin{eqnarray}
\pi(x,y) & = & \left\{{\begin{array}{ll}
0 & \mbox{iff $x=00$ or $y=00$} \\
\mbox{$1 \over 9$} & \mbox{iff $x\neq 00$ and $y \neq 00$}
\end{array}
}\right.
\end{eqnarray}
It is easily verified that for this distribution, 
every {\em deterministic\/} protocol with only 
two bits of communication will have a correctness ratio of at most
${7 \over 9}$. Using Theorem 3.20 of \cite{KN}, this shows that every
possible
randomized protocol with the same amount of communication will 
have a success ratio of at most ${7 \over 9}$.
(It can also be shown that this ${7 \over 9}$ bound is tight but we 
will omit that proof here.) 
This implies that in order to reach the requested ration of ${4 \over 5}$, 
at least three bits of communication are required if we are not allowed 
to use any prior entanglement.

\subsection{Two Qubits Suffice Without Prior Entanglement}

A similar result also holds for qubit protocols without prior 
entanglement \cite{Yao93}. 
This can be seen by the fact that after Alice
applied the rotation $A_{x_1x_2}$ and measured her qubit $Q_A$
with the result $m_A=0$, she knows the state of Bob's qubit $Q_B$ exactly.
It is therefore also possible to envision a protocol where
the parties assume the measurement outcome $m_A=0$ (this can be
done without loss of generality), and for which Alice simply sends this
qubit $Q_B$ to Bob, after which Bob finishes the protocol in the same
way as prescribed by the `prior entanglement'-protocol. The protocol has 
thus become as follows.

First Alice and Bob decide by a random joint coin flip who is going
to be the sender and the receiver in protocol. (Again we assume here that 
Alice is the sender.) Next, Alice (the sender) sends a qubit 
$\ket{Q_{x_1 x_2}}$ (according to the input string $x_1 x_2$ of Alice and 
the table 
\ref{Qtable}) to the receiver Bob who continues the protocol.
\begin{eqnarray} \label{Qtable}
\begin{array}{lcl}
\ket{Q_{00}} =  \sqm{2 \over 5} \ket{0} 
                   -\imag \sqm{3 \over 5} \ket{1} & ~ &
\ket{Q_{01}} = \sqm{4 \over 5}\ket{0} + 
 \left({\sqm{3 \over 16} + \imag \sqm{1 \over 80}}\right) \ket{1}\\
 & & \\
\ket{Q_{10}} = \sqm{4 \over 5} \ket{0} +
 \left({-\sqm{3 \over 16} + \imag \sqm{1 \over 80}}\right) \ket{1} & ~ & 
\ket{Q_{11}} = \sqm{1 \over 5}\ket{0} 
                   -\imag \sqm{4 \over 5}\ket{1}
\end{array}
\end{eqnarray}
If Bob has the input string $y_1 y_2 = 00$, he concludes the protocol
by sending a zero bit to Alice. In the other case, Bob applies the rotation
$B_{y_1y_2}$ to the received qubit, measures the qubit in the standard 
basis, and sends this measurement outcome to Alice as the answer of
the protocol. By doing so, the same correctness-probability of $4 \over 5$ 
is reached for the \IP\ function with two qubits of communication, 
whereas the classical setting requires 3 bits of communication as shown 
above. 
\eee

%
%

\section*{Acknowledgments}

\noindent We would like to thank Gilles Brassard, Harry Buhrman, Peter
H\o yer, and Tal Mor for their comments about this research.
R.C. would like to thank the Laboratoire d'Informatique Th\'eorique et
Quantique, Universit\'e de Montr\'eal for their gracious hospitality
while this research was initiated. M.N. thanks the Office of Naval
Research (Grant No.\ N00014-93-1-0116).

\section*{Appendix: Capacity Results for Communication Using Qubits}

In this appendix, we present results about the quantum 
resources required to transmit $n$ classical bits between two 
parties when two-way communication is available.
These results are used in the main text in the proof of the lower 
bound on the communication complexity of the inner product function, 
and may also be of independent interest.\ee


\loud{Theorem 2:}{\sl Suppose that Alice possesses $n$ bits of 
information, and wants to convey this information to Bob.
Suppose that Alice and Bob possess no prior entanglement but 
qubit communication in either direction is allowed.
Let $n_{AB}$ be the number of qubits Alice sends to Bob, and 
$n_{BA}$ the number of qubits Bob sends to Alice ($n_{AB}$ and $n_{BA}$ 
are natural numbers).
Then, Bob can acquire the $n$ bits if and only if the following 
inequalities are satisfied: 
\begin{eqnarray}
n_{AB} & \geq & \lceil n/2 \rceil \label{app1}\\ 
n_{AB}+n_{BA} & \geq & n \label{app2}.
\end{eqnarray}
More generally, Bob can acquire $m$ bits of mutual information 
with respect to Alice's $n$ bits if and only if the above equations hold 
with $m$ substituted for $n$.}\ee

Note that Theorem 1 follows from Theorem 2 because, if the communication 
from Bob to Alice is not counted then this can be used to set up an 
arbitrary entanglement at no cost.

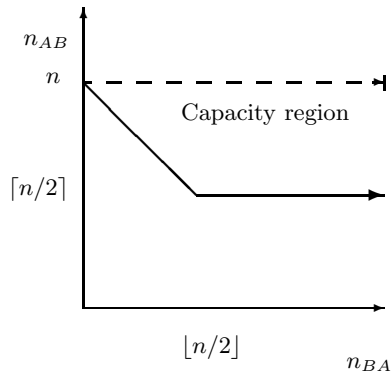
\begin{figure}[htbp]
\setlength{\unitlength}{1cm}
\begin{center}
\begin{picture}(4.5,5)
\put(1,1){\vector(0,1){4}}
\put(1,1){\vector(1,0){4}}
\put(0.2,4.5){$n_{AB}$}
\put(4.5,0.2){$n_{BA}$}
\put(0.5,4){$n$}
\put(0,2.5){$\lceil n/2 \rceil$}
\put(2.3,0.4){$\lfloor n/2 \rfloor $}
\thicklines
\put(1,4){\line(1,-1){1.5}}
\put(2.5,2.5){\vector(1,0){2.5}}
\linethickness{0.2mm}
\put(1,4){\dashbox{0.2}(4,0){}}
\thinlines
\put(4.84,4){\vector(1,0){0.16}}
\put(2.3,3.5){Capacity region}
\end{picture}
\end{center}
\vspace{-7mm}
\caption{Capacity region to send $n$ bits from Alice to Bob.
$n_{AB}$ is the number of qubits Alice sends to Bob,
and $n_{BA}$ is the number of qubits Bob sends to Alice. The
dashed line indicates the bottom of the classical capacity region.
\label{fig: capacity}}
\end{figure}

Graphically, the capacity region for the above communication problem
is shown in Fig.~\ref{fig: capacity}. Note the difference with
the classical result for communication with bits, where the capacity region
is given by the equation $n_{AB} \geq n$; that is, classically, 
communication from Bob to Alice does not help.

\loud{Proof of Theorem 2:}
The sufficiency of Eqns.~(\ref{app1}) and (\ref{app2}) follows from the 
superdense coding technique \cite{BW92}.
The nontrivial case is where $n_{AB} < n$.
Bob prepares $n - n_{AB} \le n_{BA}$ EPR pairs and sends one qubit of each 
pair to Alice, who can use them in conjunction with sending 
$n - n_{AB} \le n_{AB}$ qubits to Bob to transmit $2(n - n_{AB})$ 
bits to Bob.
Alice uses her remaining allotment of $2n_{AB} - n$ qubits to transmit the 
remaining $2n_{AB} - n$ bits in the obvious way.

The proof that Eqns.~(\ref{app1}) and (\ref{app2}) are necessary follows 
from an application of Holevo's Theorem \cite{Holevo73}, which we now review.
Suppose that a classical information source produces a random variable $X$.
Depending on the value, $x$, of $X$, a quantum state with density operator 
$\rho_x$ is prepared.
Suppose that a measurement is made on this quantum state in an effort to 
determine the value of $X$.
This measurement results in an outcome $Y$.
Holevo's theorem states that the mutual information $I(X:Y)$ between $X$ 
and $Y$ is bounded by the {\em Holevo bound} \cite{Holevo73}
\begin{eqnarray}
I(X:Y) \leq S(\rho) - \sum_x p_x S(\rho_x),
\end{eqnarray}
where $p_x$ are the probabilities associated with the different values of 
$X$, $\rho = \sum_x p_x \rho_x$, and $S$ is the von Neumann entropy function.
The quantity on the right hand side of the Holevo bound is known as the 
{\em Holevo chi quantity}, 
$\chi(\rho_{{}_X}) = S(\rho)-\sum_x p_x S(\rho_x)$.


Let $X$ be Alice's $n$ bits of information, which is uniformly distributed 
over $\01^n$.
Without loss of generality, it can be assumed that the protocol between 
Alice and Bob is of the following form.
For any value $(x_1,\ldots,x_n)$ of $X$, Alice begins with a set of 
qubits in state $\ket{x_1,\ldots,x_n}\ket{0,\ldots,0}$ and Bob begins 
with a set of qubits in state $\ket{0,\ldots,0}$.
The protocol first consists of a sequence of steps, where at each step 
one of the following processes takes place.
\begin{enumerate}
\item
Alice performs a unitary operation on the qubits in her possession.
\item
Bob performs a unitary operation on the qubits in his possession.
\item
Alice sends a qubit to Bob.
\item
Bob sends a qubit to Alice.
\end{enumerate}
After these steps, Bob performs a measurement on the qubits in his 
possession, which has outcome $Y$.
(Note that one might imagine that the initial states could be mixed and 
that measurements could be performed in addition to unitary operations; 
however, these processes can be simulated using standard techniques 
involving ancilla qubits.)

Let $\rho_i^X$ be the density operator of the set of qubits that are 
in Bob's possession after $i$ steps have been executed.
Due to Holevo's Theorem, it suffices to upper bound the final value 
of $\chi(\rho_i^X)$---which is also bounded above by $S(\rho_i)$.
We consider the evolution of $\chi(\rho_i^X)$ and $S(\rho_i)$.
Initially, $\chi(\rho_0^X) = S(\rho_0) = 0$, since Bob begins in a 
state independent of $X$.
Now, consider how $\chi(\rho_i^X)$ and $S(\rho_i)$ change for each of the 
four processes above.
\begin{enumerate}
\item
This does not affect $\rho_i^X$ and hence has no effect on 
$\chi(\rho_i^X)$ or $S(\rho_i)$.
\item
It is easy to verify that $\chi$ and $S$ are invariant under unitary 
transformations, so this does not affect $\chi(\rho_i^X)$ and $S(\rho_i)$ 
either.
\item
Let $B$ denote Bob's qubits after $i$ steps and $Q$ denote the qubit that 
Alice sends to Bob at the $i+1^{\mbox{\scriptsize st}}$ step.
By the subadditivity inequality and the fact that, for a single qubit $Q$, 
$S(Q) \leq 1$, $S(BQ) \le S(B)+S(Q) \le S(B) + 1$.
Also, by the Araki-Lieb inequality \cite{AL70}, 
$S(BQ) \ge S(B) - S(Q) \geq S(B) - 1$.
It follows that $S(\rho_{i+1}) \le S(\rho_i) + 1$ and 
\begin{eqnarray}
\chi(\rho_{i+1}^X) 
& = & S(\rho_{i+1}) - \sum_{x \in \01^n} p_x S(\rho_{i+1}^x) \nonumber \\
& \le & (S(\rho_i)+1) - \sum_{x \in \01^n} p_x (S(\rho_i^x)-1) \nonumber \\
& = & \chi(\rho_i^X) + 2.
\end{eqnarray}
\item
In this case, $\rho_{i+1}^X$ is $\rho_i^X$ with one qubit traced out.
It is known that tracing out a subsystem of any quantum system does not 
increase $\chi$ \cite{Schumacher96}, so 
$\chi(\rho_{i+1}^X) \le \chi(\rho_i^X)$.
Note also that $S(\rho_{i+1}) \le S(\rho_i) + 1$ for this process, by 
the Araki-Lieb inequality \cite{AL70}.
\end{enumerate}

Now, since $\chi(\rho_i^X)$ can only increase when Alice sends a qubit 
to Bob and by at most 2, Eq.~(\ref{app1}) follows.
Also, since $S(\rho_i)$ can only increase when one party sends a qubit to 
the other and by at most 1, Eq.~(\ref{app2}) follows.
This completes the proof of Theorem 2.

\end{document}